\documentclass[aps,pre,twocolumn,showpacs,floatfix]{revtex4}

\usepackage{amsmath}
\usepackage{amsfonts}
\usepackage{amssymb}
\usepackage{graphicx}

\begin{document}

\title{Inertial effects in B{\"u}ttiker-Landauer Motor and Refrigerator
at the Overdamped Limit}

\author{Ronald Benjamin}
\author{Ryoichi Kawai}
\email{kawai@uab.edu}

\affiliation{Department of Physics, University of Alabama at Birmingham,
Birmingham, Alabama 35294, USA}

\date{\today}

\begin{abstract}
We investigate the energetics of a Brownian motor driven by position
dependent temperature, commonly known as the B{\"u}ttiker-Landauer
motor. Overdamped models ($M=0$) predict that the motor can attain
Carnot efficiency. However, the overdamped limit ($M\rightarrow 0$),
contradicts the previous prediction due to the kinetic energy
contribution to the heat transfer. Using molecular dynamics simulation
and numerical solution of the inertial Langevin equation, we confirm
that the motor can never achieve Carnot efficiency and verify that the
heat flow via kinetic energy diverges as $M^{-1/2}$ in the overdamped
limit. The reciprocal process of the motor, namely the
B{\"u}ttiker-Landauer refrigerator is also examined. In this case, the
overdamped approach succeeds in predicting the heat transfer only when
there is no temperature gradient. Its found that the Onsager symmetry
between the motor and refrigerator does not suffer from the singular
behavior of the kinetic energy contribution.
\end{abstract}

\pacs{05.40.-a, 05.40.Jc, 05.10.Gg, 05.70.Ln}

\maketitle

\section{Introduction}
\label{sec:introduction}

Since the industrial revolution, thermodynamics has been a guiding
principle for the development of new technology.  We are now entering
the era of nanotechnology where we can construct and manipulate
nanoscale objects as we desire. Realization of nanoscale machines is
within our reach. However, we still have to overcome various issues. As
the size of a system approaches to that of molecules, thermal
fluctuations begin to play a significant role. It would be rather
difficult to operate a nanomachine against strong thermal fluctuations.
Instead, the nanomachine must be able to work harmoniously or even
collaboratively with the fluctuations. In order to design such machines,
we need to understand thermodynamics of small systems taking into
account large thermal fluctuations. Furthermore, the molecular machinery
in biological systems such as motor proteins are similarly subject to
large thermal fluctuations. Thermodynamics, at the macromolecular level,
is also essential in the investigation of such biological machines.

Unfortunately, since it was originally developed for macroscopic systems
where fluctuations are negligible, standard thermodynamics is often
powerless in cases where fluctuations dominate.  We often resort to
stochastic approaches such as Fokker-Planck equation or Langevin
equation. Despite the fact that these approaches have been successfully
used for many years, the relation between thermodynamics and stochastic
methods is not well established. It was only 10 years ago when a general
theory of energetics such as heat within the stochastic regime
(stochastic energetics)~\cite{sekimoto97,sekimoto98,sekimoto} was
developed.  The validity of the theory needs to be systematically tested
with experiment or first principles simulation.  

When traditional thermodynamics was developed, Carnot engine played a
key role as an idealized model. Similarly,  Brownian
motors~\cite{reimann02} have been basic working models for systems
dominated by thermal fluctuations. In particular, autonomous thermal
engines such as the Feynman-Smoluchowski (FS) motor~\cite{feynman} and
B{\"u}ttiker-Landauer (BL) motor~\cite{buttiker87,landauer88}, unlike
other Brownian motors, do not require time-dependent external influence.
 These motors are driven solely by thermal fluctuations and their
motility disappears when the motors become macroscopic in size.

There is no difficulty in the investigation of their motility using
standard stochastic approaches. However, it was not straightforward to
investigate the thermodynamics of these systems. In his celebrated
textbook~\cite{feynman}, Feynman attempted to investigate the
thermodynamics of the FS motor and concluded that it can reach  Carnot
efficiency.  Yet, he overlooked the effect of fluctuations and later it
was shown that the heat transfer between two heat reservoirs never
ceases even when the motor moves quasistatically and thus it is not
possible to reach the Carnot
efficiency~\cite{sekimoto97,parrondo96,hondou98}. Even when the average
velocity of the motor is zero, fluctuations around the mean value can
transport some energy.  It is this fluctuation that transports heat.
Recently, a simpler model of the FS motor was developed and their result
confirms the presence of such heat transfer~\cite{vandenbroeck04}.
Similar heat transfer was investigated in the problems of adiabatic
piston~\cite{kestemont00} and shared piston~\cite{vandenbroeck01}. The
reverse process of the FS motor, namely the FS refrigerator has also
been studied using various
models~\cite{jarzynski99,vandenbroeck06,nakagawa06}.

The BL motor is just an overdamped Brownian particle in a periodic
potential field subject to spatially inhomogeneous temperature. When
temperature changes across a potential barrier, the Brownian particle
jumps over the barrier more often from the hot side to the cold side
than the other way~\cite{buttiker87,landauer88,bier96}. Therefore,
Brownian particles move in one direction on average.  Whereas the FS
motor is simultaneously in contact with two heat baths, the BL motor
moves from one heat bath to another by itself.  Therefore, the BL motor
is a different class of Brownian motor from the FS motor. In order for
the BL motor to operate continuously, it must be thermalized with the
local environment before entering the next heat bath. Thus, it works
better in the overdamped limit but fails in the underdamped
limit~\cite{blanter98}.

Like the FS motor, there is no difficulty in explaining the motility of
the BL motor. However, the thermodynamics of this system is not
straightforward. Intuitively we expect non-vanishing heat transfer even
when the average current is zero since the Brownian particles can move
back and forth between the hot and cold regions by thermal
fluctuations~\cite{derenyi99,hondou00}.  Despite this anticipation, some
previous investigation claimed that the BL motor can reach the Carnot
efficiency~\cite{matsuo00,asfaw04+05+07} and under certain conditions
can act as a refrigerator, attaining the corresponding Carnot
coefficient of performance~\cite{asfaw04+05+07}. It turns out, however,
that the overdamped Langevin approach fails to predict such heat
transfer for the BL motor while it worked fine for the FS
motor~\cite{sekimoto97,hondou98}. It appeared that even when the system
is in the overdamped regime, inertial mass ($M$) apparently plays a
critical role in certain thermodynamic processes.  In fact, heat
evaluated by assuming $M=0$ at the beginning does not agree with the
result obtained by taking the limit $M \rightarrow 0$ at the end. This
singularity has been phenomenologically
predicted~\cite{derenyi99,hondou00} but not yet experimentally tested.
Due to the lack of experimental confirmation, this issue is still a
subject of debate~\cite{asfaw04+05+07,ai05+06}. 

In this paper, we investigate the BL motor and its reciprocal process,
the BL refrigerator, using stochastic energetics based on Langevin
approach, with and without inertial effects and compare the results with
molecular dynamics simulation.  Our main objective is to check the
failure of the overdamped Langevin  method and the validity of the
inertial Langevin method by comparing the results with molecular
dynamics simulation. The overdamped method appears to fail for the BL
motor involving inhomogeneous temperature but works fine for the BL
refrigerator operated with homogenous temperature.  On the other hand,
these two processes are related to one another through  Onsager's
symmetry, which is not obvious for a system with inhomogeneous
temperature, prompting us to investigate the validity of the Onsager
symmetry in the overdamped regime~\cite{vankampen91}. We will check the
validity numerically, taking into account inertial effects.

In the next section, we introduce a concrete model of BL
motor/refrigerator which can be investigated by both Langevin and
molecular dynamics simulations. In
section~\ref{sec:heuristic_discussion}, a brief heuristic discussion
based on the overdamped Langevin equation is given. In
section~\ref{sec:results}  the results of the overdamped approach are
compared with numerical simulations of the Langevin equation taking
inertia into account and the molecular dynamics simulation. Further
discussion and conclusion follow in the final section.

\section{The model and methods}

\subsection{The model}

\begin{figure}
\includegraphics[width=3.3in]{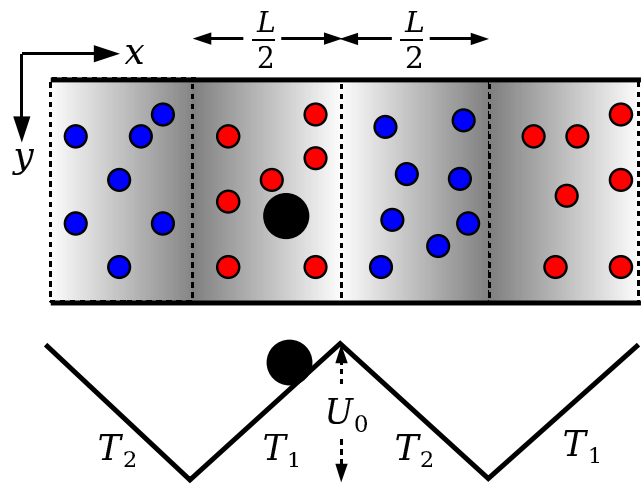}
\caption{\label{fig:model}(Color online) Two rectangular reservoirs
filled with gas particles at temperatures $T_{1}$ and $T_{2}$ are
alternately connected. Gas particles (red and blue particles) are
confined in the cell and only Brownian particles (large black
particle) can move through the walls. Brownian particles are subjected
to a piece-wise linear potential as shown in the bottom.}
\end{figure}

We consider a chain of two-dimensional cells aligned in the $x$
direction as illustrated in Fig.~\ref{fig:model}. Each cell has a width
of $L/2$ and is filled with $N$ gas particles confined in the cell they
belong to. No direct heat exchange between cells through the walls is
permitted.  The gas particles in a cell act as a heat reservoir with its
own temperature, independent of the temperature in other cells. Brownian
particles of mass $M$ are placed in the cells. Unlike the gas
particles, the Brownian particles are allowed to move freely from one
cell to another through the walls and they are also subject to a
potential field $U(x)$, which is periodic in the $x$ direction with a
periodicity $L$ and constant in the $y$ direction. For simplicity, we
use a piecewise-linear potential:
\begin{equation}
U(x) = 
\begin{cases}
\frac{2 U_{0}}{L} x & \text{for  $0 < x \leq \frac{L}{2}$}, \\
\frac{2 U_{0}}{L} (L-x) & \text{for  $\frac{L}{2} < x \leq L $},
\end{cases}
\label{eq:U}
\end{equation}
where $U_{0}$ is the potential height.  The $x$ coordinate is chosen
such that the location of potential maxima/minima coincide with the cell
boundaries as shown in Fig.~\ref{fig:model}.  In addition to the
periodic potential, a constant external force $F$ is exerted on the
Brownian particles.  We further assume that temperature is periodic with
the same periodicity as the potential and piecewise constant:
\begin{equation}
T(x) =
\begin{cases}
T_{1} & \text{for  $ 0 < x \leq \frac{L}{2} $},\\
T_{2} & \text{for  $ \frac{L}{2} < x \leq L $}.
\end{cases}
\label{eq:T}
\end{equation}
Temperature is measured in energy unit ($k_{B}=1$).

For technical simplicity, we consider only two cells (cell 1
for $0 \leq x < L/2$ and cell 2 for $L/2 \leq x < L$) with a
periodic boundary condition instead of infinitely long chains.

\subsection{Langevin approaches}

Molecular Dynamics (MD) simulation is a useful tool to investigate this
kind of model. However, a simpler mathematical model is desirable
since MD simulation is computationally quite demanding. Apart from
long-time fluid dynamical effects in two dimensions, the motion of the
Brownian particle in the $x$ direction can be investigated by the
one-dimensional Langevin equation:
\begin{equation}
\begin{gathered}
\dot{x} = v, \\
M\dot{v}=-\gamma(x) v- U^\prime(x) + F + \sqrt{2 \gamma(x) T(x)}
\xi(t),
\end{gathered}
\label{eq:Langevin}
\end{equation}
where $x$ and $v$ are position and velocity of the Brownian
particle and $\xi(t)$ is a standard Gaussian white noise:
\begin{equation}
\langle \xi(t)\rangle=0, \quad \langle \xi(t)\xi(s)\rangle
=\delta (t-s).
\end{equation}
Here, and later on, an overdot refers to derivative taken with respect
to time and a prime means derivative taken with respect to space. The
position-dependent friction coefficient $\gamma(x)$ is assumed to be
periodic and piece-wise constant in the same way as temperature:
\begin{equation}
\gamma(x) =
\begin{cases}
\gamma_{1} & \text{for  $ 0 < x \leq \frac{L}{2} $},\\
\gamma_{2} & \text{for  $ \frac{L}{2} < x \leq L $}.
\end{cases}
\label{eq:gammaposition}
\end{equation}

It has been shown that the Langevin Eq.~(\ref{eq:Langevin}) correctly
predicts the behavior of Brownian particles. However, due to its
mathematical difficulty, further approximation is often used. When the
relaxation time $\tau=M/\gamma$ is much smaller than a mechanical time
scale $t_0=\sqrt{ML^2/U_0}$,  the inertial term in Langevin
equation~(\ref{eq:Langevin}) is usually neglected. Setting $M=0$ in
Eq.~(\ref{eq:Langevin}) we obtain a popular overdamped Langevin
equation,
\begin{equation}
\gamma(x) \dot{x} = -U^\prime(x) + F + 
\sqrt{ 2 \gamma(x) T(x)} \xi(t)\, .
\label{eq:wrong_overdamped_Langevin}
\end{equation}
While this equation is widely used in many different subject areas, we
must be careful with $M=0$ since both $\tau$ and $t_0$ are zero at the
same time.  Strictly speaking, the overdamped condition $\tau \ll t_0$
should be satisfied only in the sense of the limit $M \rightarrow 0$.

Further complication arises when temperature or friction coefficient
depends on the position.  Simple omission of the inertial term does not
lead to the correct overdamped Langevin equation. One can also study 
stochastic processes in the overdamped limit by the equivalent
Fokker-Planck equation. There is, however, no universal Fokker-Planck
equation that describes a system with non-uniform temperature. Van
Kampen~\cite{vankampen88} found that a particular form of Fokker-Planck
equation depends on the details of each system.  For our model
(overdamped Brownian particle subject to inhomogeneous temperature) the
appropriate Fokker-Plank equation is given by,
\begin{equation}
\frac{\partial P(x,t)}{\partial t}= 
\frac{\partial}{\partial x} \left \{
\frac{1}{\gamma(x)}
\left [ (U^\prime
(x)-F) + \frac{\partial}{\partial x} T(x)
\right ] P(x,t) \right \}.
\label{eq:Fokker-Planck}
\end{equation}
While solving Eq.~(\ref{eq:Fokker-Planck}) for the piecewise constant
temperature~(\ref{eq:T}), the proper boundary conditions are
$T_{1}P(\delta)=T_{2}P(L-\delta)$ and
$T_{1}P(L/2-\delta)=T_{2}P(L/2+\delta)$, where $\delta$ is
infinitesimally small~\cite{boundary}.

Corresponding to the Fokker-Planck equation~(\ref{eq:Fokker-Planck}),
the correct form of the overdamped Langevin equation, in the
Stratonovich interpretation, is
\begin{equation}
\begin{split}
\gamma(x) \dot{x} &= -U^{\prime}(x)+F
 +\sqrt{2 T(x)\gamma(x)}\xi(t) \\
& -\frac{1}{2{\gamma(x)}}\frac{d}{dx}[T(x)\gamma(x)]
\end{split}
\label{eq:overdamped_Langevin}
\end{equation}
as derived in \cite{sancho92,jayannavar95}. In this paper, we shall call
Eq.~(\ref{eq:overdamped_Langevin}) the \emph{overdamped} Langevin
equation and refer to Eq.~(\ref{eq:Langevin}) as the \emph{inertial}
Langevin equation. 

As an indicator of overdamping, we introduce a dimensionless frictional
coefficient:
\begin{equation}
 \hat{\gamma} = \frac{\gamma t_0}{M} = \frac{\gamma L}{\sqrt{MU_0}}\;.
\label{eq:dimentionless_gamma}
\end{equation}
When $\hat{\gamma} >> 1$, the system is in the overdamped regime. 

\subsection{Stochastic energetics}

We investigate the thermodynamic behavior of the BL system using
stochastic energetics introduced by
Sekimoto~\cite{sekimoto97,sekimoto98,sekimoto}. Heat flux from the gas
particles in the $i$-th cell to the Brownian particles is defined as
\begin{equation}
 \dot{Q}_i = \left \langle \left (-\gamma_i \dot{x} + \sqrt{2 \gamma_i
T_i} 
\xi(t) \right )  \dot{x} \right \rangle_i
\label{eq:heat_definition}
\end{equation}
where $\langle \cdots \rangle_i$ indicates ensemble average taken while
the Brownian particles are located in the $i$-th cell. Using the
inertial Langevin equation~(\ref{eq:Langevin}),
Eq.~(\ref{eq:heat_definition}) can be evaluated  in three terms:
\begin{align}
 \dot{Q}_i &=  \frac{M}{2} \frac{d}{dt}\left \langle \dot{x}^2 \right
\rangle_i
+ \left \langle U^\prime (x)  \dot{x} \right \rangle_i 
- F \left \langle \dot{x} \right \rangle_i \notag \\
&=\dot{Q}^\text{\sc KE}_i + \dot{Q}^\text{\sc PE}_i + \dot{Q}^\text{\sc
J}_i
\label{eq:heat_inertial_Langevin}
\end{align}
where the first two terms on the rhs are the kinetic energy and potential
energy contribution to the heat flux, respectively, and the last term is
the Joule heat. 
 
Work done on a Brownian particle by the external force in each cell
is given by,
\begin{equation}
 \dot{W}_i = F \langle \dot{x} \rangle_i\;.
\label{eq:work}
\end{equation}
In the steady state, the net energy flux to the Brownian particles must
be zero, and thus the energy gained by the Brownian particles in cell 1
must be cancelled by the energy loss in cell 2.  Therefore, heat flux
from cell 1 to cell 2 via the Brownian particles is defined as
\begin{equation}
 \dot{Q}_{1 \rightarrow 2} = \dot{Q}_1 + \dot{W}_1 =  \dot{Q}^\text{\sc KE}_1 +
\dot{Q}^\text{\sc PE}_1\;.
\label{eq:heat_transport}
\end{equation}

\begin{figure}
\begin{center}
\includegraphics[width=3.3in]{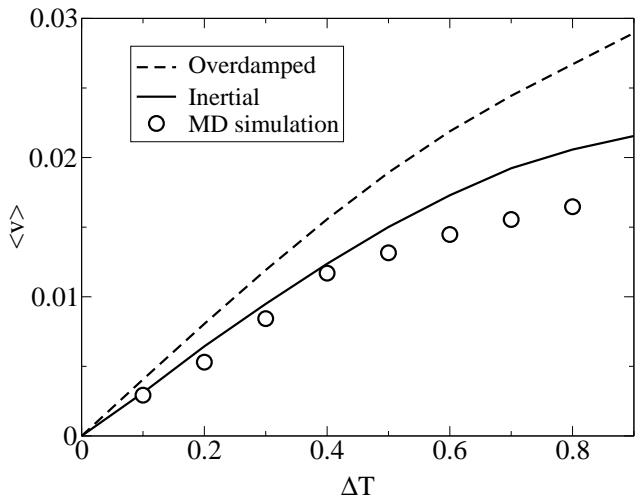}
\end{center}
\caption{\label{fig:v-dT}Mean velocity as a function of the temperature
difference $\Delta T = T_1 - T_2$ between the two cells. The average
temperature of the whole system is fixed to
$T_{av}=(T_{1}+T_{2})/2=0.5$. Solid and dashed lines correspond to data
obtained from the inertial Langevin equation (\ref{eq:Langevin}) and the
overdamped Langevin equation (\ref{eq:overdamped_Langevin})
respectively, while circles correspond to MD simulation. The parameter
values are $M/m=4.0$, $\sigma_{B}=4.0$ and  $F=0$. The dimensionless
frictional coefficients vary from
$\hat{\gamma}_{1}=\hat{\gamma}_{2}=17.7$ at $\Delta T=0$ to
$\hat{\gamma}_{1}=24.4$ and $\hat{\gamma}_{2}=5.6$ at $\Delta T=0.9$. As
$\hat{\gamma}_{2}$ becomes less overdamped with the increase of $\Delta
T$, the overdamped model deviates from the inertial Langevin equation.}
\end{figure}

\subsection{Molecular Dynamics simulation}

In order to check the validity of the Langevin approach, we
performed intensive molecular dynamics simulation. In our MD simulation
the heat bath consists of two-dimensional hard disks of mass $m$ and
diameter $\sigma$. The Brownian particle is of mass $M$ and diameter
$\sigma_{B}$. Inclusion of external forces significantly reduces the
numerical advantage of hard disk MD simulation. However our algorithm
(see Appendix~\ref{sec:appendix_MD}) is fast enough to realize a
sufficient number of trajectories for ensemble average. 

In order to compare the results of Langevin approaches with that of MD
simulation, we need to find the corresponding frictional coefficient. We
use an analytical expression obtained for an ideal gas~\cite{meurs04}:
\begin{equation}
 \gamma = \sigma_B \rho \sqrt{ 2 \pi m T}\;,
\label{eq:gamma}
\end{equation}
where $\rho$ is the density of the gas.  Note that this ideal $\gamma$
does not depend on $M$.  Alternately, we could use $\gamma$ measured in
MD simulation. Infact, we can get a better agreement between MD
simulation and Langevin approaches if the measured values are used.
However, the measured values depend on $M$, hindering the real mass
dependency of the Langevin equation.  Therfore, we use the theoretical
frictional coefficient~(\ref{eq:gamma}) in the Langevin
equations~(\ref{eq:Langevin}) and~(\ref{eq:overdamped_Langevin}).

In all our MD simulations we use $N=1000$ gas particles ($m=1$,
$\sigma=1$) in each cell, with the size of each cell being $250 \times
400$ so that the density $\rho=0.01$. The spatial period of the
piece-wise linear potential is $L=500$ and the barrier height
$U_{0}=1.0$. As the temperature of the reservoirs change during the
course of the simulation there is no well defined stationary state.
However, all our simulations were carried out using a sufficiently large
number of gas particles so that the system remains in a quasi-steady
state for a sufficiently long period of time, enabling us to measure the
various physical quantities properly.

\section{Heuristic discussion with the overdamped model}
\label{sec:heuristic_discussion}

In this section we first review the known properties of the BL motor and
its reciprocal process BL refrigerator, in the overdamped limit. When
the temperature of the cells are different ($T_1 > T_2$), the Brownian
particles in the high temperature cell can reach a higher potential
energy region than those in the low temperature cell.  Hence, the
Brownian particles tend to move from the hot cell to the cold cell over
the potential barrier.  At the other cell boundary the potential is
minimum and both cold and hot Brownian particles can easily cross to the
other side.  Therfore, the Brownian particles flow in the positive $x$
direction on average even in the absence of external force ($F=0$). When
an external load is applied ($F < 0$), the Brownian particles can do
work against it as a motor.

\begin{figure}
\begin{center}
\includegraphics[width=3.3in]{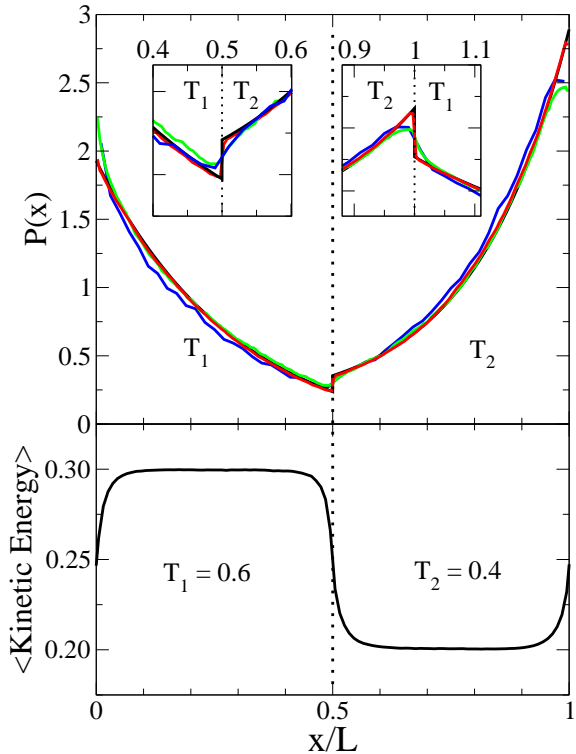}
\end{center}
\caption{\label{fig:px-ke}(Color online) Upper panel: Steady state
spatial distribution $P(x)$ of Brownian particles as obtained from 
Fokker-Planck equation (\ref{eq:Fokker-Planck}) [black line], overdamped
Langevin equation~(\ref{eq:overdamped_Langevin}) [red line], inertial
Langevin equation~(\ref{eq:Langevin}) [green line] and MD simulation
[blue line]. The parameter values  are $M/m=5.0$, $\sigma_{B}=6.0$,
$T_1=0.6$, $T_2=0.4$ and $F=0$. $\hat{\gamma}_1=26.0$ and
$\hat{\gamma}_2=21.3$ correspond to the overdamped regime. The two
insets show the details of $P(x)$ near the temperature boundaries at
$x/L=0.5$ and $x/L=1.0$. Lower panel: The locally averaged kinetic
energy of the Brownian particle along the $x$ axis, obtained from the
inertial Langevin equation~(\ref{eq:Langevin}).}
\end{figure}

Asymmetry in the spatial distribution of the Brownian particles due to
the temperature difference is the main driving of this motor.  The
overdamped Langevin equation~(\ref{eq:overdamped_Langevin}) or
Fokker-Planck equation~(\ref{eq:Fokker-Planck}) provide an analytical
expression of the spatial distribution, from which we obtain the average
velocity of the motor (See
Appendix~\ref{sec:appendix_overdamped_Langevin})
\begin{widetext}
\begin{equation}
\langle v \rangle = 
\frac{2 \sinh \left (\phi_1 +\phi_2 \right )}
{\left ( \frac{\gamma_1}{\phi_1} - \frac{\gamma_2}{\phi_2} \right )
\left (\frac{1}{f_1} - \frac{1}{f_2} \right )
 \sinh\phi_1 \sinh\phi_2  
 - \left ( \frac{\gamma_1}{f_1} + \frac{\gamma_2}{f_2} \right )
\sinh\left (
\phi_1 + \phi_2 \right ) }\;,
\label{eq:velocity_overdamped}
\end{equation}
\end{widetext}
where $f_{1,2}=-F \pm 2U_0/L$ and $\phi_i = -f_i L /(4T_i)$. In the
absence of external force ($F=0$),
Equation~(\ref{eq:velocity_overdamped}) is always positive for
$T_{1}>T_{2}$ and hence the Brownian particles move in the positive
direction as expected. In fact, particle distribution and velocity in
the overdamped regime agree reasonably well with the results obtained
from numerical simulation of the inertial Langevin equation and
molecular dynamics simulation (see Figs.~\ref{fig:v-dT}
and~\ref{fig:px-ke}), suggesting the validity of the overdamped Langevin
equation. 

Now, we consider the thermodynamic efficiency of the BL motor. Based on
the successful prediction of particle distribution and velocity it is
natural to use the overdamped Langevin approach for other quantities
such as efficiency.  Applying the overdamped Langevin
equation~(\ref{eq:overdamped_Langevin}) to the definition of
heat~(\ref{eq:heat_definition}), the heat flux from the heat bath to the
Brownian particles in cell $1$ is given by
\begin{equation}
 \dot{Q}_1 = \left \langle
\frac{1}{2\gamma(x)}\frac{d}{dx}[\gamma(x)T(x)]  \dot{x} \right
\rangle_1\ + \left \langle U^\prime(x)  \dot{x} \right \rangle_1 -
F\left
\langle 
\dot{x} \right \rangle_1\;.
\label{eq:heat_overdamped_Langevin}
\end{equation}
Comparing this equation with Eq.~(\ref{eq:heat_inertial_Langevin}), the
first term in the rhs of Eq.~(\ref{eq:heat_overdamped_Langevin})
corresponds to the kinetic energy contribution $\dot{Q}^\text{\sc
KE}_i$. However, due to periodic boundary condition and
Eq.~(\ref{eq:gamma}), this kinetic energy contribution
vanishes~\cite{matsuo00} and we obtain $\dot{Q}_1 = (2 U_0/L-F) \langle
\dot{x} \rangle_1$. Hence, we find the efficiency of the motor
\begin{equation}
 \eta = \frac{-\dot{W} }{\dot{Q}_1} = \frac{-2F}{2
U_0/L-F}\;,
\end{equation}
where $\dot{W}=F\langle v \rangle$. When the motor is in a stalled
state, ($\phi_1+\phi_2=0$), we can show that the efficiency reaches the
Carnot efficiency $\eta_\text{\sc C} = 1-T_2/T_1$.

This result is puzzling since at the moment a Brownian particle enters
the cold bath it is carrying $T_1/2$ of kinetic energy. When it is
thermalized with the cold bath, $Q^\text{\sc KE}_{1 \rightarrow 2}
\propto (T_1-T_2)/2$ of heat dissipates into the cold bath.  This heat
dissipation takes place whenever the Brownian particle crosses a
temperature boundary and it is irreversible. Due to diffusion the
crossing occurs even when the average velocity vanishes, and this
irreversible heat due to thermal fluctuation of Brownian particles never
ceases as long as there is a steep temperature gradient.
Der\'{e}nyi-Astumian~\cite{derenyi99} and
Hondou-Sekimoto~\cite{hondou00} have pointed out that the kinetic energy
contribution is the dominant channel of heat transfer between two cells
and thus the efficiency of the BL motor cannot reach  the Carnot
efficiency. This situation is similar to the case of the FS
motor~\cite{parrondo96,sekimoto97}.

An interesting question is why the overdamped Langevin approach failed
for the BL motor whereas it worked fine for the FS motor.
Der\'{e}nyi-Astumian~\cite{derenyi99} and also
Hondou-Sekimoto~\cite{hondou00} phenomenologically answered this
question. Consider the thermal velocity $v_{th} = \sqrt{T/M}$. When a
Brownian particle crosses the temperature boundary, it will not be
immediately thermalized and there is a narrow region $\ell_{th} = v_{th}
\tau = \sqrt{TM}/\gamma$   where the average kinetic energy of the
Brownian particle does not coincide with the temperature of the
reservoir (see lower panel of Fig.~\ref{fig:px-ke}). In the overdamped
limit ($M \rightarrow 0$) this region vanishes, justifying the use of
the overdamped approach.  However, the effective temperature gradient is
$|T_1-T_2|/\ell_{th} \propto M^{-1/2}$.  Hence, the heat current is
proportional to $M^{-1/2}$ and diverges as $M \rightarrow 0$. This
singularity implies that the overdamped regime that assumes $M=0$ is not
equivalent to the overdamped limit $M \rightarrow 0$.   Apparently, the
FS motor does not have this problem~\cite{sekimoto97}.  One of our
objectives is to verify this singular behavior by numerical simulations.

Next, we turn to the BL refrigerator. We assume $T_1=T_2=T$ and apply a
weak external force $0<F<2U_0/L$ to the Brownian particle. Due to the
external force, the Brownian particle drifts in the direction of $F$ and
its average velocity in the overdamped limit can be obtained from
Eq.~(\ref{eq:velocity_overdamped}). Unlike the motor case, the
temperature is uniform throughout the system and thus the kinetic energy
does not contribute significantly to the heat exchange between the
cells.  Again using the overdamped model, the heat flux absorbed by the
Brownian particle from the $i$-th reservoir is obtained from
Eq.~(\ref{eq:heat_overdamped_Langevin}) as
\begin{equation}
\dot{Q}_1=+\frac{2U_0}{L}\langle \dot{x} \rangle_1 -F \langle \dot{x}
\rangle_1, \quad 
\dot{Q}_2=-\frac{2U_0}{L}\langle \dot{x} \rangle_2 -F \langle \dot{x}
\rangle_2\,,
\label{eq:heat_fridge_overdamped}
\end{equation}
where the first term is the potential energy contribution $\dot{Q}^\text{\sc
PE}_i$ and the second term the Joule heat $\dot{Q}^\text{\sc J}_i$.
Since $\dot{Q}^\text{\sc PE}_i \propto F$ and $\dot{Q}^\text{\sc J}_i
\propto F^2$ we can have positive $\dot{Q}_1$ for sufficiently small
$F$, indicating that cell 1 is refrigerated.   Since there is no
singular behavior due to temperature change, the overdamped model may be
good enough.  However, as soon as refrigeration induces a temperature
difference between the cells, the heat transfer due to the kinetic
energy reduces the power of refrigeration.  Our second objective is to
investigate whether the overdamped model is sufficient and if the
refrigeration can be sustained against the heat leak due to the kinetic
energy. 

The BL motor and refrigerator are a result of cross effect between the
external force $F$ and the temperature difference $\Delta T$.  Assuming
they are small, we can make a connection between the motor and
refrigerator using  linear irreversible thermodynamics expressed by
\begin{equation}
\label{eq:linear_response}
\begin{gathered}
\langle v \rangle  = L_{11}\frac{F}{T} + L_{12}\frac{\triangle
T}{T^{2}},
\\
\dot{Q}_{1 \rightarrow 2}  = L_{21}\frac{F}{T} + L_{22}\frac{\triangle
T}{T^{2}},\\
\end{gathered}
\end{equation}
where the transport coefficients are defined as
\begin{equation}
\begin{split}
 L_{11} &= T \frac{\partial \langle v \rangle}{\partial
F}, \qquad
 L_{12} = T^{2} \frac{\partial \langle v \rangle}{\partial
\Delta T}, \\
 L_{21} &= T \frac{\partial \dot{Q}_{1 \rightarrow 2}}{\partial
F}, \quad
 L_{22} = T^2  \frac{\partial \dot{Q}_{1 \rightarrow 2}}{\partial
\Delta T}\,.
\end{split}
\label{eq:transport_coefficients}
\end{equation}
Here, the partial derivatives are evaluated at $\Delta T=0$  and $F=0$.
In the overdamped case we can obtain all Onsager coefficients from
Eqs.~(\ref{eq:heat_transport}),
({\ref{eq:velocity_overdamped}), and
({\ref{eq:heat_overdamped_Langevin}), as
\begin{equation}
L_{11} = \frac{U_0^2}{T}\xi, \,
L_{22} = \frac{U_0^4}{TL^2}\xi, \,
L_{12}=L_{21}=\frac{{U_{0}}^{3}}{T L}\xi,
\label{eq:transport_coefficients_overdamped}
\end{equation}
where $\xi = [4 \gamma \sinh^2(U_0/2T)]^{-1}$.  These coefficients
support the Onsager symmetry $L_{12}=L_{21}$.
Moreover, Eq.~(\ref{eq:transport_coefficients_overdamped})
indicates $L_{11}L_{22}-L_{12} L_{21}=0$, implying no entropy
production~\cite{degroot} and thus the motor operates at the Carnot
efficiency. Hence the overdamped model again erroneously predicts the
highest possible efficiency. The singular behavior of $\dot{Q}_{1
\rightarrow 2}$ at $M=0$ casts some doubt on the Onsager symmetry since
the two limits $M \rightarrow 0$ and $\Delta T \rightarrow 0$ do not
commute. Our third objective is to numerically investigate the validity
of linear irreversible theory and the Onsager symmetry at the overdamped
limit.

\section{Results}
\label{sec:results}

\subsection{Motor}
\label{sec:results_motor}

\begin{figure}
\begin{center}
\leavevmode
\includegraphics[width=3.3in]{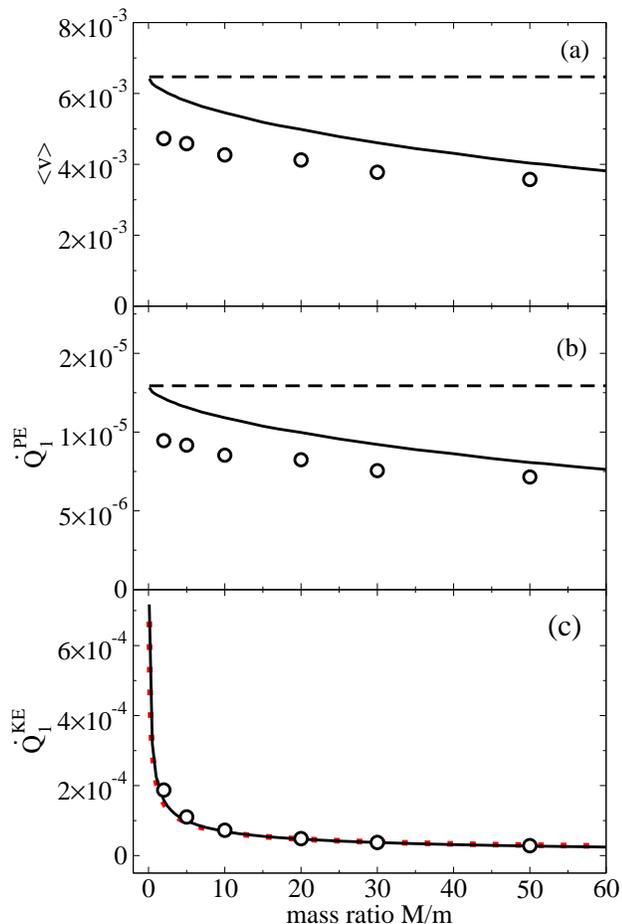}
\caption{\label{fig:v+Q_M}(Color online) Mean velocity (a), heat flux
due to potential energy (b), and heat flux due to kinetic energy (c) as
a function of the mass ratio $M/m$. Circles, dashed lines and solid
lines correspond to MD simulation, overdamped model and the inertial
Langevin equation, respectively. The parameter values are $T_1=0.6$ and
$T_2=0.4$, $\sigma_{B}=5.0$ and $F=0.0$. As the mass ratio changes from
$M/m=0.1$ to $M/m=60$, $\hat{\gamma}_1$ varies from $48.5$ to $6.3$,
and $\hat{\gamma}_{2}$ varies from $39.6$ to $5.1$. In (c) the dotted
line indicates $0.000215 {(M/m)}^{-1/2}$, an empirical fit to the result
of inertial Langevin equation.}
\end{center}
\end{figure}

First we show in Fig.~\ref{fig:px-ke} that the numerical solution of
overdamped Langevin equation~(\ref{eq:overdamped_Langevin}) agrees
perfectly with the solution of the corresponding Fokker-Planck
equation~(\ref{eq:Fokker-Planck}), confirming that the extra term in
Eq.~(\ref{eq:overdamped_Langevin}) is necessary. Furthermore, the
results of overdamped Langevin equation are in good agreement with the
numerical results from inertial Langevin equation (\ref{eq:Langevin})
and molecular dynamics simulation, except for narrow regions at the
temperature boundaries (see the insets in Fig.~\ref{fig:px-ke}).
Although it is in general small, this error in the overdamped regime
should not be overlooked.

As we discussed in section \ref{sec:heuristic_discussion}, the overdamped
Langevin method assumes that the Brownian particles are locally in
equilibrium with the the thermal reservoir. Therefore, the average
kinetic energy of the Brownian particles is assumed to be the same as
the local temperature of the heat bath. This assumption, however, fails
near the temperature boundaries since the Brownian particles entering
from one temperature region to another are not immediately thermalized
with the new environment even when the mass is very small as
Fig.~\ref{fig:px-ke} indicates.  The size of the transition region is
the thermalization length $\ell_{\text{th}}$.  In the overdamped limit
($M \rightarrow 0$), indeed the transition region vanishes but only
slowly as $\sqrt{M}$. 

It turns out that such a subtle error in the overdamped Langevin method
does not lead to large errors in certain quantities obtained from it. 
Figure \ref{fig:v-dT} shows that the overdamped Langevin equation
predicts the average velocity as well as the inertial Langevin equation
does. Due to this success we feel safe to use the overdamped approach to
compute other quantities. However, the calculation of the heat is a
different story. As we discussed in the previous section, the overdamped
approach predicts that the heat transfer between the cells is only from
potential energy contribution. However, Hondou-Sekimoto \cite{hondou00}
pointed out that the narrow transition region plays a dominant role in
heat transfer and the overdamped Langevin equation fails in the
investigation of heat calculation. The inertial Langevin approach
concludes that the kinetic energy contribution is actually dominant.

\begin{figure}
\includegraphics[width=3.3in] {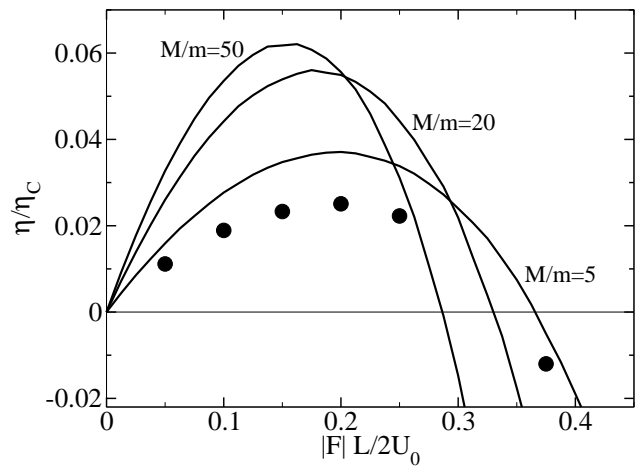}
\caption{\label{fig:motor_efficiency} The efficiency of the motor vs the
magnitude of the external load, for mass ratios $M/m=5.0,20.0$ and
$50.0$. Solid lines correspond to numerical solution of the inertial
Langevin equation and circles represent MD simulation (for $M/m=5.0$
only). Other parameter values are $T_1=0.7$, $T_2=0.3$, and
$\sigma_{B}=5.0$. The dimensionless friction coefficients are
$\hat{\gamma}_{1}=23.4$, $\hat{\gamma}_{2}=15.3$ for $M/m=5.0$;
$\hat{\gamma}_{1}=11.7$, $\hat{\gamma}_{2}=7.7$ for $M/m=20.0$; and
$\hat{\gamma}_{1}=7.4$, $\hat{\gamma}_{2}=4.9$ for $M/m=50.0$.}
\end{figure}

Using scaling arguments, Der\'{e}nyi-Astumian \cite{derenyi99} and
Hondou-Sekimoto \cite{hondou00} predicted that the kinetic energy
contribution diverges as $M^{-1/2}$ at $M=0$. Figure~\ref{fig:v+Q_M}
clearly shows such a singularity in  good agreement with their theory
and confirms $\dot{Q}^\text{\sc KE}_1 \gg \dot{Q}^\text{\sc PE}_1$.
We also find from Fig.~\ref{fig:v+Q_M} that result obtained from  the
inertial Langevin equation approaches the MD simulation result as the
mass ratio $M/m$ increases. Even for small $M/m$, the inertial Langevin
equation predicts the correct order of magnitude of the motor velocity
and the heat flows, though, in general, the Langevin approach is not
applicable in such a case. The motor velocity and heat flow via
potential energy predicted by the overdamped model, deviate from the
inertial Langevin equation and MD simulation result as the mass ratio
increases because the inertial effect becomes too large to be ignored.
The good agreement between the  inertial Langevin equation and the
molecular dynamics simulation, shows that Sekimoto's definition of
stochastic energetics predicts the heat correctly even when temperature
is spatially inhomogeneous.

In Fig.~\ref{fig:motor_efficiency}, we plot the efficiency of the motor
normalized by Carnot efficiency $\eta_\text{C}$, as a function of the
external load $F$ normalized by the magnitude of the force exerted by
the periodic potential energy. In contradiction to the overdamped model,
the efficiency is far below the Carnot limit.  While this result was
phenomenologically argued for in \cite{derenyi99,hondou00}, here we
confirmed with molecular dynamics simulation and numerical solution of
inertial Langevin equation that the kinetic energy contribution greatly
reduces the efficiency. Even when the motor is operated at the
quasistatic limit with a stall force, the irreversible heat transfer via
kinetic energy persists and thus the Carnot limit is unattainable. 
Furthermore, the divergence of $\dot{Q}^\text{\sc KE}_1$ at $M=0$
diminishes the efficiency of the motor to zero, in the overdamped limit.

\begin{figure}
\includegraphics[width=3.3in] {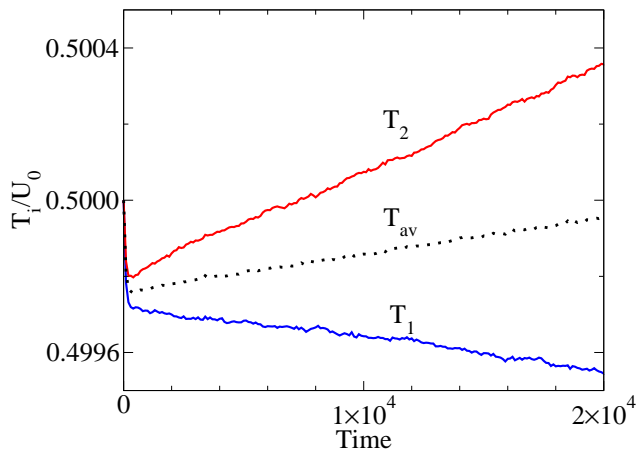}
\caption{\label{fig:cooling}(Color online) Temperature of cells 1 and 2
and the average temperature $T_{av}={(T_{1}+T_{2})}/{2}$ as a function
of time obtained from MD simulation. Initial temperatures of cells 1 and
2 are $T_{1}=T_{2}=T=0.5$ and an external force ${FL}/{(2U_{0})}=0.5$ is
applied to the Brownian particle in the positive direction. The other
parameter values are ${M}/{m}=20$ and $\sigma_{B}=8.0$,  leading to a
dimensionless frictional coefficient
$\hat{\gamma}_1=\hat{\gamma}_2=15.9$ (overdamped regime). Temperature
drops at the beginning because the Brownian particles are initially at
rest.}
\end{figure}

\subsection{Refrigerator}
\label{sec:refrigerator}

In the refrigerator mode, two cells initially have the same temperature
and the Brownian particles are driven by an external force $F$. In
Fig.~\ref{fig:cooling},  MD simulation illustrates the cooling of cell 1
at the expense of the heating of cell 2. Note however, that the average
temperature $T_{av}$ increases with time as Joule heat is dissipated in
both the cells. In the Langevin approach, we cannot see such temperature
changes since the temperature needs to be kept constant.  However, it
allows us to investigate heat transfer between the cells.
Figure~\ref{fig:Q-F_fridge} shows the components of heat from cell 1 to
the Brownian particles as a function of $F$.  The potential energy
contribution increases linearly with $F$ whereas Joule heat decreases
($\dot{Q}_{1}^{J} < 0$) as $F^2$ [see
Eq.~(\ref{eq:heat_fridge_overdamped})]. For sufficiently small $F$, the
potential energy contribution wins and hence the Brownian particles
extract heat from cell 1 and dump it in cell 2. There is an optimal $F$
at which the cooling effect is maximum. The agreement between the
overdamped and inertial Langevin equations and with the molecular
dynamics simulation is good for $\dot{Q}^\text{\sc PE}_1$ and
$\dot{Q}^\text{\sc J}_1$.

\begin{figure}
\includegraphics[width=3.3in]{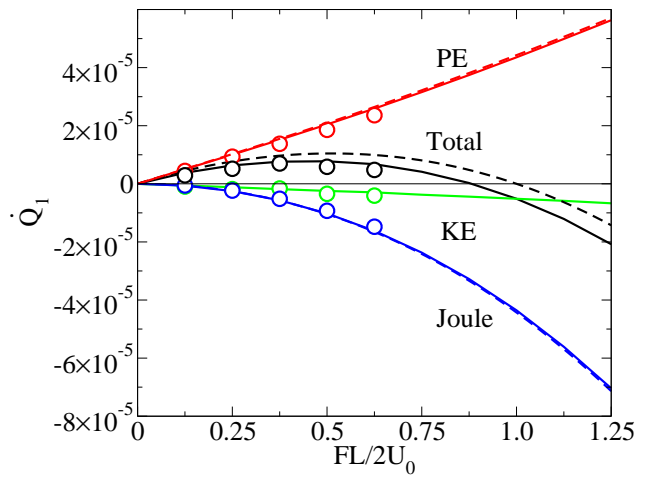}
\caption{\label{fig:Q-F_fridge}(Color online) The various components of
$\dot{Q}_{1}$ (from top to bottom the heat flows are $\dot{Q}^\text{\sc
PE}_{1}$, $\dot{Q}_{1}$, $\dot{Q}^\text{\sc KE}_{1}$ and
$\dot{Q}^\text{\sc J}_{1}$ respectively) of the refrigerator as a
function of the external force. The parameter values are $M/m=20.0$,
$\sigma_{B}=8.0$ and $T_1=T_2=0.5$ so that
$\hat{\gamma}_1=\hat{\gamma}_2=15.9$ (overdamped). Circles represent MD
data, dashed lines correspond to the overdamped Langevin equation and 
solid lines represent the inertial Langevin equation.}
\end{figure}

Since temperature is uniform, we don't expect significant kinetic energy
contribution.  However in Fig.~\ref{fig:Q-F_fridge}, we found that there
is a small amount of kinetic energy contribution which is absent in the
overdamped approach.  In the overdamped regime we assume that the change
in potential energy immediately dissipates into the reservoir. However,
with a finite mass, the potential energy is first converted to kinetic
energy which dissipates at a later time.  For example, when a Brownian
particle slides down the potential slope near a cell boundary, the potential
energy change is transported to the next cell as kinetic energy where it is
dissipated as heat. The amount of ``unthermalized'' kinetic energy the
Brownian particles acquire from the potential energy is approximately,
\begin{equation}
 \dot{Q}_1^\text{\sc KE} \approx -U'(x) \ell_\text{th} \frac{\langle
\dot{x} \rangle}{L} = -\frac{2U_0\langle \dot{x}
\rangle}{\gamma L^2} \sqrt{MT}.
 \label{eq:Q_KE_fridge}
\end{equation}
which is linear with $F$ through $\langle\dot{x}\rangle$ in agreement
with the result of the inertial Langevin equation (see
Fig.~\ref{fig:Q-F_fridge}). Both this
kinetic energy contribution and the potential energy contribution are in
proportion to $F$, yet the magnitude of $\dot{Q}_1^\text{\sc KE}$ is
negligibly smaller than that of $\dot{Q}_1^\text{\sc PE}$ at the
overdamped limit. Therefore, the refrigeration is still possible.
Figure~\ref{fig:Q-M_fridge} shows that the PE contribution approaches
the overdamped model as $M \rightarrow 0$. The KE contribution
approaches to zero as $(M/m)^{1/2}$, in good agreement with the
phenomenological prediction (\ref{eq:Q_KE_fridge}). Figures 
\ref{fig:Q-F_fridge} and \ref{fig:Q-M_fridge} both show
that the overdamped model is in good agreement with the inertial
Langevin equation as well as MD simulation in predicting the velocity
and heat flows.

\begin{figure}
\includegraphics[width=3.3in]{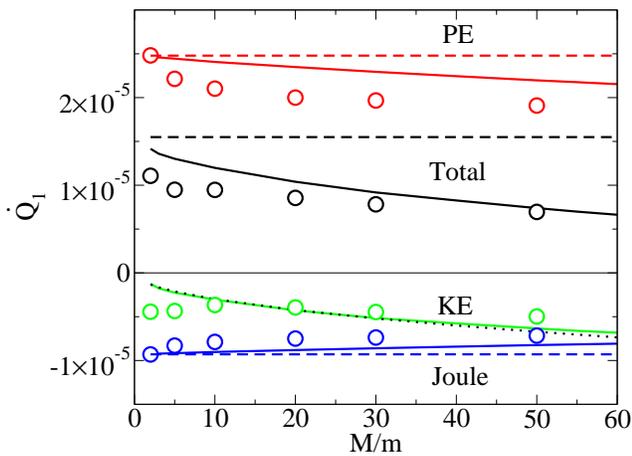}
\caption{\label{fig:Q-M_fridge}(Color online) The various components of
$\dot{Q}_{1}$ of the refrigerator as a function of the mass. The
parameter values are $\sigma_{B}=5.0$, $T_1=T_2=0.5$ and
${FL}/{(2U_{0})}=0.375$. The dimensionless friction coefficients
($\hat{\gamma}_1=\hat{\gamma}_2$) vary from $31.3$ at $M/m=2.0$
(overdamped) to $5.7$ at $M/m=60.0$ (weakly damped). Refer to
Fig.~\ref{fig:Q-F_fridge} for an explantion of the various data lines
and symbols. The dotted  line represents a fit ($\propto \sqrt{M/m}$) to
the kinetic energy contribution according to
Eq.~(\ref{eq:Q_KE_fridge}).}
\end{figure}

The overdamped Langevin equation seems to work well for the refrigerator. 
However, the cooling effect creates a temperature difference between the two
reservoirs and the overdamped approach again fails.  In turn, the temperature
difference induces a thermodynamic force opposing the external force $F$,
through the Brownian motor mechanism. As the temperature difference increases,
the motor and refrigerator effects eventually cancel each other, so that the
average velocity of the Brownian particle becomes zero and the cooling ceases. 

\subsection{Onsager symmetry}

As discussed in section~{\ref{sec:heuristic_discussion}, we need to
evaluate the transport coefficients (\ref{eq:transport_coefficients})
with a certain care. Since the heat transfer diverges when the limit $M
\rightarrow 0$ is taken before $\Delta T \rightarrow 0$, we first
evaluate the coefficients at the finite mass numerically from the
response curves like Figs.~\ref{fig:v-dT} and ~\ref{fig:Q-F_fridge}. 
Then, we reduce the mass toward $M=0$.
Figure~\ref{fig:transport_coefficients-mass} shows that as $M$
decreases, the product of off-diagonal coefficients approaches to the
result of overdamped Langevin equation obtained from
Eq.~(\ref{eq:transport_coefficients_overdamped}). However, the product
of diagonal coefficients diverges as $(M/m)^{-1/2}$, reflecting the
divergence of $\dot{Q}^\text{\sc KE}$. As expected, the off-diagonal
coefficients do not suffer from the singular inertial effect.
Furthermore, Fig.~\ref{fig:transport_coefficients-mass} shows that
$L_{11}L_{22} \gg L_{12}L_{21}$ for all values of $M/m$, indicating
large entropy production consistent with the very low efficiency.
Finally, Fig.~\ref{fig:onsager-symmetry} numerically verifies the
Onsager symmetry for all masses within the accuracy of simulation and
that the overdamped model (\ref{eq:transport_coefficients_overdamped})
predicts the correct overdamped limit.

\begin{figure}
\begin{center}
\includegraphics[width=3.3in]{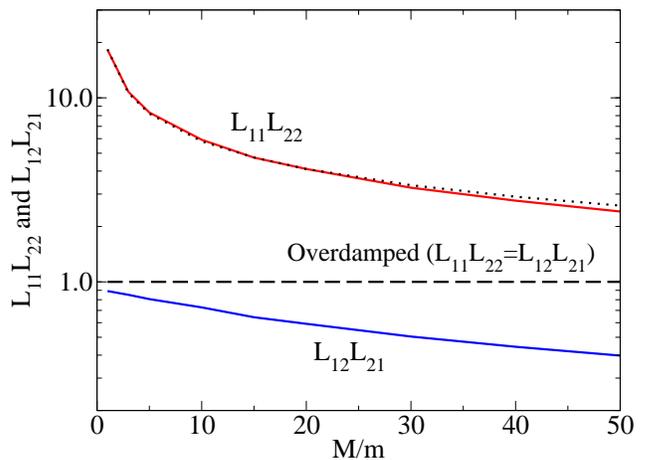}
\caption{\label{fig:transport_coefficients-mass}(Color online) The
product of the Onsager coefficients $L_{11} L_{22}$ (red curve) and
$L_{12} L_{21}$ (blue curve) as a function of $M/m$  obtained from the
inertial Langevin equation. Data has been normalized with respect to the
overdamped value. The dashed line represents Onsager coefficients
obtained from overdamped results. The parameters for obtaining $L_{12}$
and $L_{22}$ were $\Delta T=0.025$ and $F=0$, while $L_{11}$ and
$L_{21}$ were obtained at $\Delta T=0$ and ${FL}/{(2U_{0})}=0.025$,
where $T_{1}=(1/2)(1+{\Delta T})$, $T_{2}=(1/2)(1-{\Delta T})$ and
$\sigma_{B}=5.0$. We use a logarithmic scale for the y-axis. The dotted
line is the phenomenological fit $18.4 {(M/m)}^{-1/2}$, to the product
$L_{11} L_{22}$.} 
\end{center}
\end{figure}

\section{Discussion}

\begin{figure}
\begin{center}
\includegraphics[width=3.0in]{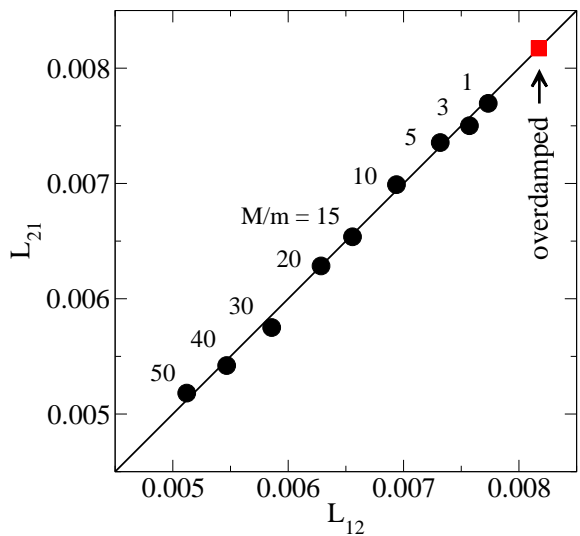}
\caption{\label{fig:onsager-symmetry}(Color online) The Onsager
reciprocity coefficient $L_{21}$ vs $L_{12}$ for various  $M/m$. The
parameter values are the same as in
Fig.~\ref{fig:transport_coefficients-mass}. The solid circles represent
data obtained from the inertial Langevin  equation. The value obtained
from the overdamped model is shown with a solid square.}
\end{center}
\end{figure}
In this paper, we investigated the thermodynamic properties of the BL
motor and refrigerator using Langevin equations and molecular dynamics
simulation.  For mechanical properties such as the velocity of the Brownian 
particles, we observed reasonable agreement between overdamped Langevin 
equation and molecular dynamics simulation.  However, the overdamped Langevin
equation failed to predict thermodynamic properties such as the heat
transfer. On the other hand, we found good agreement
between the inertial Langevin equation and molecular dynamics simulation
even in the overdamped regime.  Therefore, we conclude that the inertial
mass plays a significant role even in the overdamped limit. The main
effect of the inertial mass is the kinetic energy contribution to the
heat transfer.  We confirmed that the kinetic energy contribution is
dominant when two cells have different temperature and verified a
previous phenomenological prediction, that the irreversible kinetic
energy contribution diverges as $M^{-1/2}$ at the overdamped limit.

Recently, Van den Broeck~\cite{vandenbroeck07} investigated the
efficiency of Brownian motors using linear irreversible thermodynamics
and concluded that, in principle, Carnot efficiency can be attained. He
argues that when the load $F$ is small, the temperature difference
$\Delta T$ drives the motor in the forward direction and at the same
time the motor transfers heat from the high temperature reservoir to the
low temperature reservoir. As the load exceeds the stall force, the
motor moves backward, transferring heat against the temperature
gradient as a heat pump. In an ideal system, the direction of velocity
and heat are reversed at the same magnitude of the external force.
Hence, the thermodynamic fluxes $\langle v \rangle$ and $\dot{Q}_{1
\rightarrow 2}$ simultaneously vanish despite $F$ and $\Delta T$ not
being zero. This can only happen when $L_{11}L_{22}-L_{12}L_{21}=0$ and
thus Carnot efficiency is achieved.

In our case, the direction of $\dot{Q}^\text{\sc PE}$ coincides with the
direction of $\langle v \rangle$, and both $\dot{Q}^\text{\sc PE}$ and
$\langle v \rangle$ vanish at the stall force. However,
$\dot{Q}^\text{\sc KE}$ flows from the hot to cold reservoir regardless
of the direction of $\langle v \rangle$. Therefore, $\dot{Q}^\text{\sc
KE}$ not necessarily vanishes. In fact, it never vanishes and the Carnot
efficiency cannot be achieved in the BL system.
The efficiency of the BL motor is significantly reduced by the kinetic
energy contribution and vanishes at the overdamped limit.

D{\'e}renyi-Astumian~\cite{derenyi99} have argued that if we could
prevent the re-crossing of the Brownian particle over the temperature
boundary, such irreversible heat transfer could be arbitrarily reduced
and thus the efficiency would be greately improved. They proposed to
place a gate at each temperature boundary which prevents Brownian
particles from crossing the boundary back and forth multiple times
during a short period of time. If such a gate is possible, the motor
could reach Carnot efficiency.  However, it is not clear at present how
to construct such a gate without reducing the particle velocity.
Based on a naive consideration one might expect significant reduction of
the kinetic energy contribution by optimizing the potential profile and
the location of the temperature boundary. Unfortunately, the reduction
of heat transfer always results in the reduction of the motor velocity
as well and the gain in the efficiency is very limited~\cite{benjamin}.

Recently, Humphrey et al.~\cite{humphrey02} proposed a
quantum Brownian heat engine which achieves Carnot
efficiency. In this model, reversible particle
exchange between two cells is achieved by filtering the energy of the
particle, without violating second law.  The filter only allows 
particles having a certain energy, for which the Fermi-Dirac
distribution in the two cells coincide, to pass through. This ensures
that the particle flow does not alter the thermal distribution in 
either cell so that there is no kinetic energy contribution. Hence, the
heat engine attains Carnot efficiency in the
quasistatic limit. Whether a similar heat engine can be devised in the
classical regime is not clear.

\section{Acknowledgments}
We would like to thank Ken Sekimoto and C. Van den Broeck for useful
discussion. The computer simulation was partly carried out at the
Alabama Supercomputer Center.

\appendix
\section{Hard Disk Molecular Dynamics Simulation}
\label{sec:appendix_MD}

Our hard disk molecular dynamics simulation is based on a usual
event-driven algorithm~\cite{rapaport}. In the case of collision
between two gas particles the collision time can be obtained
analytically as given in~\cite{rapaport}. However, since the Brownian
particle is subject to forces due to the piece-wise linear potential
$U(x)$ and the external load $F$,  the calculation of collision time
between a Brownian particle and a gas particle is not straightforward.
Suppose that we know the relative position $\textbf{r}$ and relative
velocity $\textbf{v}$ of a Brownian particle with respect to a gas
particle at time $t_0$. We want to find the time $t=t_0+\epsilon$, at
which they collide. The distance between two particles at the moment of
impact is $d = (\sigma + \sigma_B)/2$ and hence
\begin{equation}
 \left | \textbf{r} + \textbf{v}\epsilon + \frac{1}{2}\textbf{a}
\epsilon^2 \right | = d
\label{eq:collision_time}
\end{equation}
where $\textbf{a}$ is a constant acceleration of the Brownian particle.
(Note that the gas particle has no acceleration.)
To find $\epsilon$, we rewrite Eq. (\ref{eq:collision_time}) as a
quartic equation for $a\ne0$
\begin{equation}
 \epsilon^4 + A \epsilon^3 + B\epsilon^2 +
C\epsilon + D = 0 \;,
\label{eq:quartic}
\end{equation}
where 
$A=4\textbf{v}\cdot\textbf{a}/a^2$,
$B=4(\textbf{a}\cdot\textbf{r}+v^2)/a^2$, 
$C=8\textbf{r}\cdot\textbf{v}/a^2$, and $E=4(r^2-d^2)/a^2$. Using a
standard
method ~\cite{hacke41}, we find four solutions
\begin{equation}
\epsilon =-\frac{A}{4} + \frac{1}{2} \left( F \pm \sqrt{ G + H }\right),
\quad
\epsilon =-\frac{A}{4} - \frac{1}{2} \left( F \pm \sqrt{ G - H }\right),
\label{eq:roots}
\end{equation}
where,
\begin{eqnarray}
 F &=& \pm \sqrt{\frac{A^2}{4}-B+x} \\
 G &=& \frac{3}{4}A^2-F^2-2B\\
 H &=& 
\begin{cases}
 \displaystyle\frac{4AB - 8C- A^3}{4F} & \text{if} \quad
F\ne 0
\medskip\\
 2 \sqrt{x^2 - 4D} & \text{if} \quad F=0
\end{cases}
\end{eqnarray}
and $x$ is a real solution of an auxiliary equation,
\begin{equation}
 x^3 - B x^2 + (AC-4D)x + 4BD-C^2-A^2D=0 \;.
\label{eq:cubic}
\end{equation}
The analytical solutions to the cubic equation~(\ref{eq:cubic}) are
given in~\cite{press}.

The correct collision time corresponds to the smallest positive real
root of Eq.~(\ref{eq:roots}), if it exists.  Despite having analytical
solutions this algorithm occasionally fails, particularly when the
acceleration $a$ is significantly small because some of the coefficients
become extremely large causing bit-off errors.  To overcome this
difficulty, we improve the accuracy of the roots by iterating the
Newton-Raphson~\cite{press} steps a few times starting from the
analytical solution to Eq.~(\ref{eq:cubic}).  We found that the present
algorithm is faster than directly solving the quartic
equation~(\ref{eq:quartic}) by the Newton-Raphson method.

\section{Analytical Expressions in Overdamped Case}
\label{sec:appendix_overdamped_Langevin}

We derive analytical expressions for steady state density
$P(x)$ and current $J(x)$.  The Fokker-Planck equation for steady
states is simply $dJ(x)/dx= 0$,  where the particle current is
defined by
\begin{equation}
\gamma(x) J(x)  = -(U^{\prime}(x)-F) P(x)
-\frac{d}{dx} [T(x) P(x)].
\label{eq:currden}
\end{equation}
In our model, the Fokker-Planck equation for the $i$-th cell is given
by
\begin{equation}
P_i^{\prime \prime}(x) + \frac{f_i}{T_i} P_i^{\prime}(x) = 0,
\quad (i=1,2),
\end{equation}
where the net force is defined as $f_{1,2}=-F \pm 2U_0/L$. We consider
only weak external forces $|F| < F_c = 2U_0/L$ so that it does not
destroy the potential barrier and hence we assume $f_i \ne 0$. Then,
we find general solutions:
\begin{align}
P_1(x) &=
C_1 \exp \left ( -\frac{f_1 x}{T_1} \right ) + D_1, \label{eq:P1}  \\
P_2(x) &=
C_2 \exp \left [ -\frac{f_2(x -L)}{T_2} \right ] + D_2, \label{eq:P2}
\end{align}
where $C_i$ and $D_i$ are the constants of integration. The
corresponding currents are $J_i = -D_i f_i /\gamma_i, \,(i=1,2)$. Since
temperature and frictional coefficient are discontinuous at the cell
boundaries, the density does not have to be continuous. However, the
current must be continuous ($J_1=J_2=J$).  From Eq.~(\ref{eq:currden}),
we find the magnitude of the density discontinuity as $T_1 P_1(L/2)=T_2
P_2(L/2)$ and $T_1 P_1(0)=T_2 P_2(L)$ \cite{boundary}. These boundary
conditions along with normalization of the density are sufficient to
determine the integral constants.  After obtaining an expression for $J$
and the integral constants, we obtain a general expression for the average
velocity [Eq. (\ref{eq:velocity_overdamped})] from a general relation
$\langle v \rangle = JL$. The particle density in each cell is obtained
from Eqs.~(\ref{eq:P1}) and (\ref{eq:P2}).


\begin{thebibliography}{99}
\bibitem{boundary}
The boundary conditions for a system with inhomogeneous temperature is
discussed in Ref.~\cite{landauer88}. Different boundary conditions,
namely $P_{1}(0)=P_{2}(L)$ and $P_{1}(L/2)=P_{2}(L/2)$ are often used in
the literature \cite{bier96,asfaw04+05+07}. However, these boundary
conditions are not consistent with the physical system under
consideration. Indeed, the solution obtained with these boundary
conditions  disagrees with the results of our molecular dynamics
simulation both qualitatively and quantatively.
\end{thebibliography}
\end{document}